\documentclass[aps,twocolumn,showpacs,superscriptaddress,amsmath]{revtex4}
\usepackage{graphics,epsf}
\usepackage{graphicx}
\newcommand{\bea}{\begin{eqnarray}}
\newcommand{\beq}{\begin{equation}}
\newcommand{\eea}{\end{eqnarray}}
\newcommand{\eeq}{\end{equation}}
\begin{document}
\title{Finite-thickness effects in ground-state transitions of 
two-electron quantum dots}
\author{R. G. Nazmitdinov}
\affiliation{Departament de F{\'\i}sica, Universitat de les Illes Balears, E-07122
Palma de Mallorca, Spain}
\affiliation{Bogoliubov Laboratory of Theoretical Physics, 
Joint Institute for Nuclear Research, 141980 Dubna, Russia}
\author{ N. S. Simonovi\'c}
\affiliation{Institute of Physics, P.O. Box 57, 11001 Belgrade,
Serbia}
\date{\today}
\begin{abstract}
Using the exactly solvable excitation spectrum of  two-electron quantum dots 
with parabolic potential, we show that the inclusion of the vertical extension  of the 
quantum dot provides a consistent description of the experimental findings
of Nishi {\it et al.} [Phys.Rev.B {\bf 75}, 121301(R) (2007)]. 
We found that the second singlet-triplet transition in the ground state 
is a vanishing function of the lateral confinement in the three-dimensional case, 
while it always persists in the two-dimensional case. 
We show that a slight decrease of the lateral confinement  
leads to a formation of the Wigner molecule at low magnetic fields.
\end{abstract}
\maketitle

Two-electron quantum dots (QDs)
have drawn a great deal of experimental and theoretical attention 
in recent years \cite{mak,kou,RM}.
Experimental data including transport measurements 
and spin oscillations in the ground state under a perpendicular magnetic field 
in two-electron QDs may be explained  as a result of the interplay between 
electron correlations, a two-dimensional (2D) lateral confinement and 
magnetic field. A 2D interpretation of experiments, however, leads to 
inconsistencies \cite{haw,kou}, providing, for example, too low values of the 
magnetic field for the first singlet-triplet (ST) transition. There is no consensus on
origin of this disagreement, since various experiments are dealing 
with different QDs. Evidently, it is important to understand 
basic sources of such inconsistencies from view point of possible
technological applications, since QDs may provide a natural realization of quantum 
bit. This problem is also related to fundamental aspects of 
strongly correlated {\it finite} systems, which are different from bulk 
and can be controlled experimentally. 

It was predicted that the ground state of N-electron QD becomes the spin polarized 
maximum density droplet (MDD) \cite{mac} at high magnetic field. For a two-electron 
QD it is expected that the MDD occurs after a first ST
transition  (see discussion in \cite{RM}). Theoretical calculations \cite{Wag} 
assert that after the first ST transition the increase of the magnetic field induces 
several ground state transitions to higher orbital-angular and spin-angular momentum states. 
This issue was addressed in  transport study of the correlated 
two-electron states up to 8T  and 10 T in a lateral \cite{ihn} and vertical 
\cite{ni,ni2} QDs, respectively. It is quite difficult to detect the structure of ground 
states after the first ST transition in a lateral QD due to a strong suppression 
of the tunnel coupling between the QD and contacts. Altering the lateral 
confinement strengths, the transitions beyond the first ST transition are 
reported in vertical QDs \cite{ni}. In fact, the variation of the confining frequency with 
{\it the same experimental setup} opens a remarkable opportunity 
in the consistent study of effects of the magnetic field on electron  
correlations. 
One of the major aspects of the present paper is to demonstrate that the experimental
results found in Ref.\onlinecite{ni} can be explained if one takes into account the
3D physical nature of the QD. We will discuss the additional 
criterion to distinguish the 2D and 3D nature experimentally and will
analyze the formation of the Wigner molecule in the 3D 
two-electron QD.

  Three  vertical QDs with different lateral confinements have been studied
in  the experiment \cite{ni}.  In all samples  clear shell structure effects
for an electron number  $N=2,6,...$ at $B=0$ T have been observed, implying 
a high rotational symmetry. 
Although there is a sufficiently small deviation from this symmetry in sample C 
(from now on in accordance with the list of Ref.\onlinecite{ni}), a complete shell
filling for two and six electrons was observed.
Such a shell structure is generally associated with
a 2D harmonic oscillator (x-y) confinement \cite{kou}.
However, it is noteworthy that a similar
shell structure is produced by a 3D axially-symmetric 
harmonic oscillator (HO) if the confinement  in the z-direction 
$\omega_z=1.5\omega_0$ is only slightly larger than the lateral confinement
($\omega_x=\omega_y=\omega_0$).  In this case  six electrons fill the lowest 
two shells with  Fock-Darwin energy levels with $n_z=0$. 
It was found  also that the lateral confinement frequency for the axially-symmetric 
QD decreases with the increase of the electron number \cite{Mel}, since the 
screening in the lateral plane becomes stronger with large electron number. 
In turn, this effectively increases the ratio $\omega_z/\omega_0$ making the dot 
to be more "two-dimensional", since the 
vertical confinement is fixed by the sample thickness. Indeed, the N-dependence 
of the effective lateral frequency is observed in \cite{ni}. All these facts 
imply that the three-dimensional nature is a prerequisite of a consistent quantitative 
analysis of small QDs with a few electrons. 

Our analysis is carried out by means of the exact diagonalization of
the Hamiltonian for two 3D interacting electrons in a perpendicular magnetic field:
\beq
\label{ham}
H = \sum_{j=1}^2 \bigg[ \frac{1}{2m^*\!}\,
\Big({\bf p}_j - \frac{e}{c} {\mathbf A}_j \Big)^{\! 2}
+ U({\mathbf r}_j) \bigg]
+ \frac{\alpha}{\vert{\mathbf r}_1 \!- {\mathbf r}_2\vert}+
H_{\it spin}.
\eeq
Here $\alpha = e^2/4\pi\varepsilon_0\varepsilon_r$ and 
$H_{\it spin}=g^*\mu_B({\bf s}_1+{\bf s}_2)·{\bf B}$ describes the Zeeman energy,
where $\mu_B=e\hbar/2m_ec$ is the Bohr magneton.
The effective mass is $m^*=0.067m_e$, the relative dielectric constant 
of a semiconductor is $\varepsilon_r=12$ and $|g^*|=0.44$ (bulk GaAs values).
For the perpendicular magnetic field we choose the vector potential with
gauge ${\mathbf A} = \frac{1}{2} {\mathbf B}
\times {\mathbf r} = \frac{1}{2}B(-y, x,0)$. The confining potential is
approximated by a 3D axially-symmetric HO $U({\mathbf
r}) = m^* [\omega_0^2\,(x^2 \!+ y^2) + \omega_z^2z^2]/2$, where
$\hbar\omega_z$ and $\hbar\omega_0$ are the energy scales of confinement in
the $z$-direction and in the $xy$-plane, respectively.

\begin{figure}[ht]
\includegraphics[height=0.22\textheight,clip]{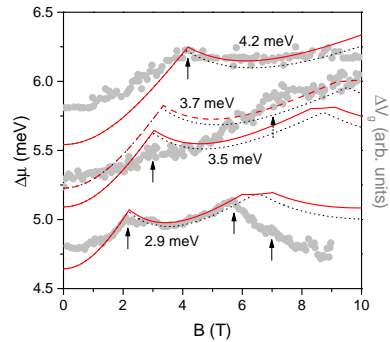}
\caption{(Color online) The magnetic dependence of the additional energy
$\Delta \mu$ in two-electron QDs with lateral confinements  
$\hbar\omega_0=4.2,3.7,3.5,2.9$ meV (the first, second and fourth values 
are experimental values for sample A, B, and C, respectively \cite{ni}). 
The confinement in the third (z) direction $\hbar\omega_z=8$ meV is fixed for all samples. 
The results for $|g^*|=0.3(0.44)$  are  connected by solid (dotted) line 
for $\hbar\omega_0=4.2,3.5,2.9$ meV and by dashed (dotted) line for 
$\hbar\omega_0=3.7$ meV.
The solid grey lines display the experimental 
spacing $\Delta V_g$ as a function of $B$. 
The arrows identify the position of experimental ground state transitions \cite{ni}.} 
\label{fig1}
\end{figure}

The evolution of the ground-state energy of a two-electron QD under the
perpendicular magnetic field can be traced
by means of the additional energy $\Delta \mu=\mu(2,B)-\mu(1,B)$, where 
$\mu(N,B)=E(N,B)-E(N-1,B)$  and $E(N,B)$ denotes the total energy of 
the QD with $N$ electrons under a magnetic field of the strength $B$ \cite{kou}.
Fitting the $B$-field dependence of the first and second Coulomb oscillation 
peak positions to the lowest Fock-Darwin energy levels 
of the 2D HO with the potential $m^*\omega_0^2r^2/2$,
Nishi {\it et al.} \cite{ni} estimated $\omega_0$
for all three samples A, B, C. Although the general
trend in the experimental data is well reproduced by the 2D calculations, 
the experimental positions of the ST transition 
points are systematically higher (see Fig.3 of Ref.\cite{ni}). 
Different  lateral confinements in the above experiment are achieved
by the variation of the electron density, without changing the sample thickness.
Using the "experimental" values for the lateral confinement and
the confinement frequency $\omega_z$ as a free parameter,
we found that the value $\hbar\omega_z=8$ meV provides the best fit for 
the positions of kinks in the additional energy 
\begin{equation}
\Delta \mu=\hbar\omega_0\varepsilon-E(1,B)+E_Z
\label{dmu}
\end{equation}
in all three samples.
Here, $\hbar\omega_0\varepsilon$ is the relative energy, 
$E(1,B)=\hbar\omega_0+\hbar\omega_z/2$ and the Zeeman energy $E_Z$ is zero for
the singlet states (we recall that
the total Hamiltonian (\ref{ham}) is separated onto the center-of-mass Hamiltonian, 
the Hamiltonian of the relative motion and the Zeeman term). 
For the sake of illustration, we display in Fig.\ref{fig1} 
the magnetic dependence of the experimental spacing between 
the first and the second Coulomb oscillation peaks 
$\Delta V_g=V_g(2)-V_g(1)$ for samples A--C, which can be transformed 
to the additional energy $\Delta \mu$ (see details in Refs.\onlinecite{ni,ni2}). 
In the $\Delta V_g-B$ plot, ground state transitions appear as upward
kinks and shoulders \cite{ni2}. It was found from the Zeeman splitting 
at high magnetic fields that $|g^*|=0.3$ \cite{ni2} 
and  we calculate the additional energy with this and the bulk values.

We nicely reproduce the experimental position of the first ST transitions
at $B=4.2,3,2.3$ T in samples A, B, and C, respectively (see Fig.\ref{fig1}). 
When the magnetic field is low, a difference between 
the calculations with different $|g^*|$ factors is negligible. 
Upon decreasing  the lateral confinement $\hbar\omega_0$ from 
sample A to sample C (the increase of the ratio $\omega_z/\omega_0$),
the Coulomb interaction becomes dominant in the
interplay between electron correlations and the confinement \cite{NSR}. 
In turn, the smaller the lateral confinement at fixed thickness is (the stronger is
the electron correlations) the smaller the value of the magnetic field is at
which the ST transitions or, in general, crossings between excited states and the
ground state may occur. 

There is no signature of the second crossing in the ground
state for sample A at large $B$ (up to $10$ T). Here, the ratio 
$\omega_z/\omega_0\approx 1.9$ and the effect of the 
third dimension is most visible: the confinement has a dominant role in the 
electron dynamics and very high magnetic field is required to observe the next 
transition in the ground state due to 
electron correlations. Thus, the MDD phase survives until very high magnetic fields 
($B \sim 10$ T). 

A second kink is observed at $B=7$ T in sample B \cite{ni}.  
Our calculations with the "experimental" lateral confinement $\hbar\omega_0=3.7$ meV 
produces the second kink at $B=9.5$ T, which is located higher than the experimental 
value. The slight decrease of the lateral frequency until $\hbar\omega_0=3.5$ 
meV shifts the second kink to $B=8.7$ T, improving the agreement with 
the experimental position of the first ST transition as well. 
In addition, the use of $|g^*|=0.3$ (instead of the bulk value) with 
the latter frequency creates a plateau, which bears resemblance to the experimental 
spacing $\Delta V_g$. However, there is no detailed information on this sample and 
we lack a full understanding of this kink. It seems
there is an additional mechanism responsible for the second kink in sample B. 

\begin{figure}[ht]
\includegraphics[height=0.22\textheight,clip]{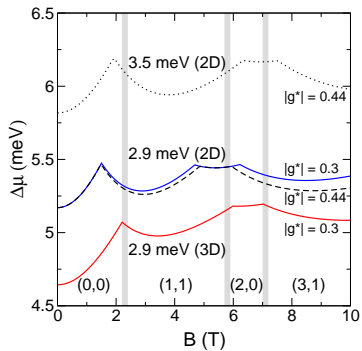}
\caption{(Color online) Magnetic field dependence of the addition energy $\Delta \mu$
for the 2D model with $\hbar\omega_0=2.9, 3.5$ meV and for the 3D model 
($\hbar\omega_0=2.9$ meV, $\hbar\omega_z=8$ meV). 
Ground states are labeled by
$(M,S)$, where $M$ and $S$ are the total orbital momentum and the total spin, 
respectively. Grey vertical lines indicate the position of the experimental 
crossings between different ground states.}
\label{fig2}
\end{figure}

The most complete experimental information is related to sample C and
we also study this sample in detail. In sample C the first experimental ST 
transition occurs at $B=2.3$ T, while the signatures of the second and the third  
ones are observed at $B\approx 5.8, 7.1$ T, respectively (see Fig.\ref{fig1}). 
The 2D calculations (with the "experimental" values $\hbar\omega_0=2.9$ meV, $|g^*|=0.44$) 
predict the first, second and third  ST crossings at lower magnetic fields: 
$B=1.7, 4.8, 5.8$ T, respectively (see Fig.\ref{fig2}).
The results can be improved to some degree with $|g^*|=0.3$.
To reproduce the data for  $\Delta \mu$ 
Nishi {\it et al.} \cite{ni}  have increased the lateral confinement
($\hbar\omega_0=3.5$ meV, $|g^*|=0.44$). As a result, the first, second and  third 
ST transitions occur at $B=2,6.3, 7.5$ T, respectively. 
Evidently, 2D calculations overestimate 
the importance of the Coulomb interaction. The increase of the lateral 
confinement  weakens simply the electron correlations in such calculations. 
In contrast, the 3D calculations reproduce quite well the positions of all crossings 
with the "experimental" lateral confinement 
$\hbar\omega_0=2.9$ meV at $B=2.3,5.8,7.1$ T (see Fig.\ref{fig2}).

\begin{figure}[ht]
\includegraphics[height=0.22\textheight,clip]{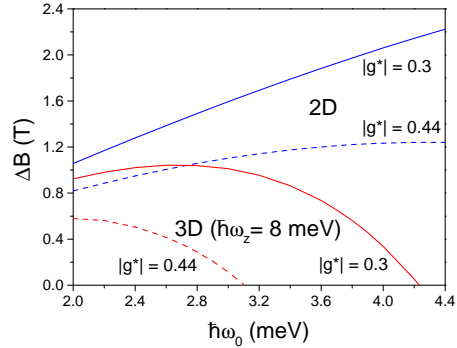}
\caption{(Color online) The interval $\Delta B$ in which the singlet
state $(2,0)$ survives as a function of the lateral confinement for 2D and 3D calculations.
The confinement in the third (z) direction 
$\hbar\omega_z=8$ meV is fixed for the 3D calculations.} 
\label{fig3}
\end{figure}

One of the questions addressed in the experiment \cite{ni} is related to 
a shoulderlike structure observed in a small range of values of the magnetic field
(see our Fig.\ref{fig1} and Fig.4 of \cite{ni}). This structure is identified as 
the second singlet state $(2,0)$ that persists till the next crossing with the 
triplet state $(3,1)$. According to Ref.\onlinecite{ni}, the ground state transition 
from the triplet $(1,1)$ state to the singlet $(2,0)$ is associated with the collapse 
of MDD state for $N=2$. Therefore, a question arises: at which conditions it would be 
possible to avoid the collapse of the MDD phase (in general, to preserve the 
spin-polarized state); i.e., at which conditions the singlet $(2,0)$ state never 
will show up in the ground state.
In fact, the collapse of the MDD depends crucially on the value of the lateral 
confinement and the dimension of the system. We found that in  the 2D consideration 
the $(2,0)$ state always exists for experimentally available lateral 
confinement (see  Fig.\ref{fig3}). 
Moreover, in this range of $\omega_0$ the 2D approach predicts the monotonic increase 
of the interval of values of the magnetic field $\Delta B$, at which the second singlet 
state survives, with the increase of the lateral confinement. 
In contrast, in the 3D calculations, the size of the interval  
is a vanishing function of the lateral confinement for a fixed thickness 
($\hbar\omega_z=8$ meV). 
It is quite desirable, however, to measure this interval 
to draw a definite conclusion and we hope it will done in future.

As discussed above, the decrease of the confinement  at fixed thickness 
increases the dominance of the electron correlations in the electron dynamics. 
Furthermore, this decrease, related to the decrease of the electron density 
\cite{Mel,ni}, creates the favorable conditions for onset of electron localization.
This localization (crystallization) in QDs is associated with the formation of the 
so-called Wigner molecule \cite{Maks}. 
In the 2D approach the crystallization is controlled by the ratio of Coulomb and 
confinement strengths  $R_W = (\alpha/l_0)/\hbar\omega_0$ 
($l_0 = (\hbar/m^*\omega_0)^{1/2}$) (cf \cite{lor}), 
which is about $R_W\sim 3$ for the QDs considered in 
experiments \cite{ni}. For a 2D two-electron QD, 
it is predicted that the Wigner molecule can be formed for $R_W\sim 200$ 
at zero magnetic field \cite{YL}, or at very high magnetic field \cite{szaf} 
(for $\hbar\omega_0\sim 3$meV and small $R_W$ 
such as in the experiments \cite{ni}).
In the 3D axially-symmetric QDs the ratio between vertical and
lateral confinements (anisotropy) may, however, affect the formation of the
Wigner molecule. 
\begin{figure}
\includegraphics[height=0.56\textheight,clip]{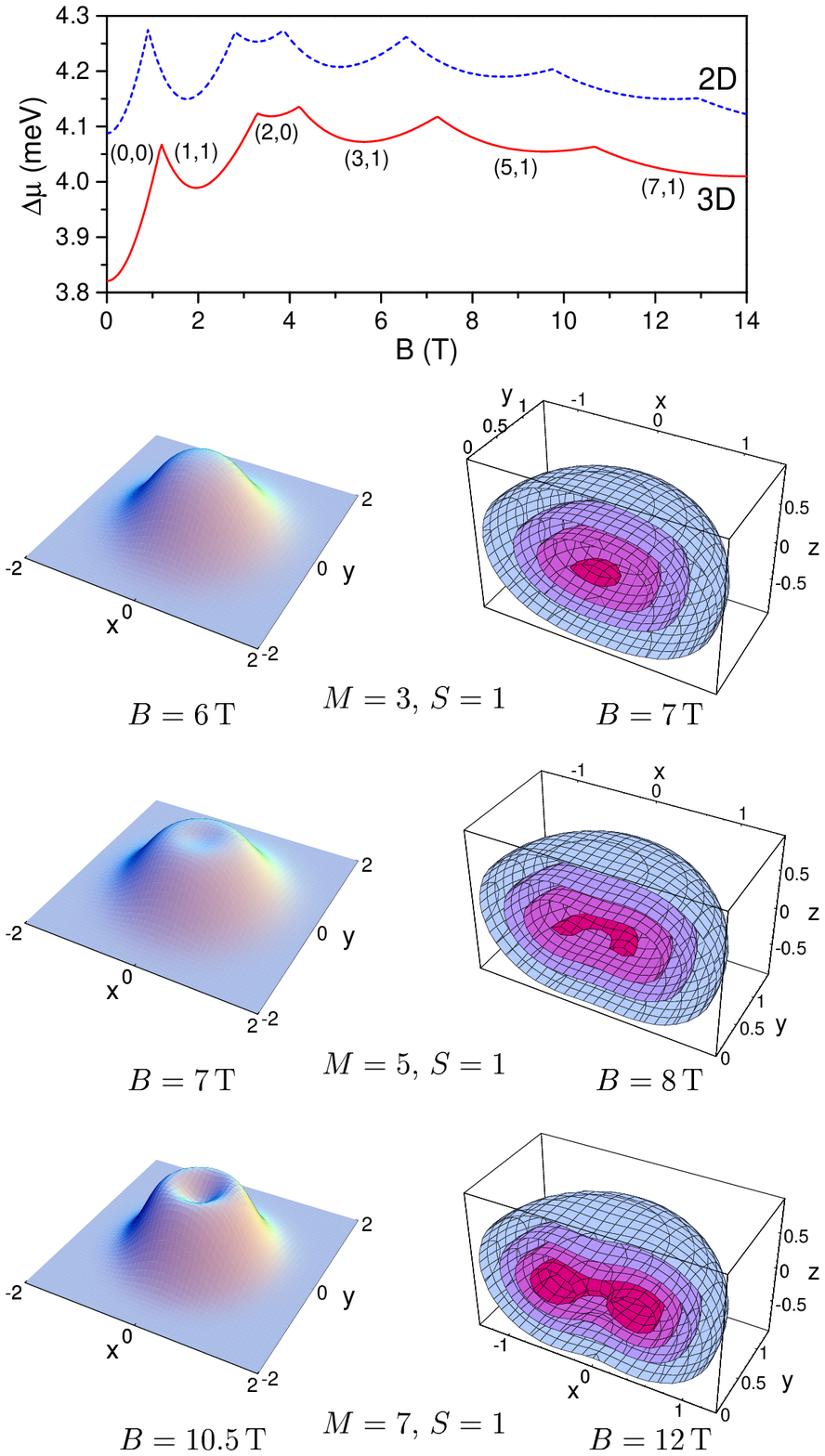}
\caption{(Color online) 
Top: the magnetic dependence of the ground state
in 2D and 3D ($\hbar\omega_z=8$ meV) approaches for a lateral confinement 
$\hbar\omega_0=2$ meV. The 2D (left) and 3D (right) electron densities are displayed 
for different ground states (M,S) at corresponding magnetic fields.
The largest 3D density grows from a central small core over a  ring to 
a  torus with the increase of the magnetic field.
}
\label{fig4}
\end{figure}
This problem can be analyzed by dint of the electron density
\beq 
n({\bf r}) = \int \left[\,|\Psi({\bf r},{\bf
r}^\prime)|^2+|\Psi({\bf r}^\prime,{\bf r})|^2 \right] {\rm d}{\bf
r}^\prime,
\label{el-density}
\eeq
when one electron is at the position $\bf r$ if the other one is located at
a position ${\bf r}^\prime$.
A criterion for the onset of the crystallization in QDs can be the
appearance of a local electron  density minimum at the center of the dot
\cite{Cref}. For 2D QDs this leads to a radial modulation in the
electron density, resulting in the formation of rings and roto-vibrational 
spectra \cite{ton}. 

Our analysis of the conditions realized in the experiments \cite{ni}
predicts very high magnetic fields ($B>12$ T) for the formation of the Wigner molecule.
However, with a slight decrease of the lateral confinement, at $\hbar\omega_0= 2$ meV  
we obtain the desired result. The 3D analysis of electron density
(see Fig.\ref{fig4}) gives an unequivocal answer that at $B > 7.25$ T the
triplet state $(5,1)$ can be associated with a formation of the Wigner
molecule. There is an evident 
difference between the 2D and 3D approaches: the 2D calculations 
predict the crystallization at lower magnetic field ($\Delta B\sim 1$ T). 
The further increase of the magnetic field leads
to the formation of a ring and a torus of  maximal density 
in 2D- and 3D-densities, respectively. Notice that if geometrical differences are disregarded,
3D evolution of the ground state can be approximately reproduced in 
2D approach with the effective charge concept \cite{NSR} (see also \cite{ihn}).

Summarizing, we have shown that the confinement in the z direction
is important ingredient for the quantitative analysis of the experimental data
for two-electron axially-symmetric vertical QDs.
In contrast to the 2D description,
the 3D approximation provides a consistent description  of
various experimental features: 
the energy spectrum for small magnetic field, the value of the magnetic
field for the first and the second singlet-triplet transitions.
We propose a criterion for the additional spectra, 
which evidently demonstrates the effect of the third dimension. According to this
criterion the singlet state $(2,0)$ is a vanishing function of the lateral confinement 
(see Fig.\ref{fig3}) in the vertical magnetic field in the two-electron 
axially-symmetric vertical QD. We found that the decrease of the lateral confinement
in the experiment \cite{ni} until $\hbar\omega_0=2$ meV  would 
lead to the formation of the Wigner molecule at $B\sim 8$ T.

\section*{Acknowledgments}
The authors are thankful to the Max-Planck-Institute for the
Physics of Complex Systems in Dresden, 
where this project was initialized. This work was partly supported 
by Project No 141029 of Ministry of
Science and Environmental Protection of Serbia, and 
by Grant No. FIS2005-02796 (MEC, Spain).

\end{document}